\documentclass[a4paper,11pt,onecolumn,UTF-8]{article}
\usepackage{fancyhdr}
\usepackage{amsmath,amsfonts,amssymb}
\usepackage{graphicx}
\usepackage{mathptmx}
\usepackage{booktabs}
\usepackage[labelfont=bf]{caption}
\usepackage{indentfirst}
\usepackage{caption}
\usepackage{enumitem}
\usepackage{subfigure}

\title{\textbf{In-phase Synchronization of Two Coupled Metronomes}}
\author{
Xuepeng Wang
\\[2pt]
{\small \textit{2016, School of Physics, Nanjing University, Nanjing 210046}}\\[6pt]
Mentor: Sihui Wang\\[2pt]
{\small \textit{School of Physics, Nanjing University, Nanjing 210046}}\\[2pt]
}
\date{}

\fancypagestyle{firststyle}
{
   \fancyhf{}
   \fancyhead[C]{Xuepeng Wang: In-phase Synchronization of Two Coupled Metronomes}
   \fancyhead[R]{\thepage}
}

\setlist{nolistsep}
\begin{document}
\maketitle
\thispagestyle{firststyle}
\setlength{\oddsidemargin}{ 1cm}
\setlength{\evensidemargin}{\oddsidemargin}
\setlength{\textwidth}{15.50cm}
\vspace{-.8cm}
\begin{center}
\parbox{\textwidth}{
\textbf{Abstract:}\quad {This paper used multi scale method and KBM method to get approximate solution of coupled Van der Pol oscillators, based on which, researchers investigated the impact several parameters have on the prerequisite of synchronization and the time it takes to synchronize quantitively. In addition, this paper has a brief introduction of the usage of Kuramoto Model in plural metronomes’ synchronization and the derivation of Van der Pol oscillator from the discrete model.} \\

}
\end{center}

\setcounter{page}{1}

\setlength{\oddsidemargin}{-.5cm}  % 3.17cm - 1 inch
\setlength{\evensidemargin}{\oddsidemargin}
\setlength{\textwidth}{17.00cm}

\section*{Introduction}
Synchronization is the phenomenon that two or more coupled systems with different initial phases interacting with each other and finally come to have same motions, which could be commonly observed between oscillators such as pendulums or metronomes. Synchronization is a fundamental theme in nonlinear phenomena and is a current popular topic of research. The earliest phenomenon of synchronization was observed by Huygens in the experiment of two coupled clock on the wall.\\

In addition, synchronization is so ubiquitous that myriad of phenomena, such as the wake-sleep cycle and the equilibrium of ecosystems, could be accredited to it. What’s more, in signal analysis and network allocation, synchronization is also an inevitable and imperative question to solve. Although the synchronization of metronomes has been explored from several perspectives, almost all previous researches were based on the widely used Van der Pol oscillator in electric engineering without necessary and precise derivation. Moreover, former researchers focus too exorbitantly on prerequisite of synchronization to have a legible insight of other paramount phenomena observed in synchronization such as synchronization time. Thus, present researchers provide a concise derivation between Van der Pol oscillator and the metronomes system and uses two different approximate method, multi scale method and KBM method, to derive the prerequisite and time of synchronization according to several parameters respectively. In addition, we also bring in Kuramoto Model into this issue to investigate the synchronization of several metronomes system. At last, we discussed the demerit of present approximate method and latent problems such approximation will cause and the prospective of the researches of synchronization.

\section{Derivation of Metronome Systems}

\subsection{Structure of metronomes.}
The parameters of a metronome are set as follow: the angle between pendulum of metronome and the vertical direction is assumed as $\theta$, and the mass of pendulum is $m$, with the distance of $r_{cm}$ between pendulum’s mass of center and the axis and $I$ as the moment of inertia of the pendulum. The total mass of other part of metronome such as its shell can be deemed as $M$.

The driving force that act on the pendulum that serves the purpose of the augment of oscillating angle and overcoming the resistance provided by friction is provided by a structure called escapement, which collide with the pendulum every time the oscillating angle $\theta$ access zero from both directions. Due to instantaneity of collision, we could assume that during the collision the angular momentum are conservative. Given that after every collision the escapement keep still, it could be elicited that the augment of angular momentum of pendulum is equal among every collision.

\subsection{Derivation of the equation.}

Thus, with the angular momentum increment to be assumed as $P_{0}$, first we write down the oscillation equation of motion of a single metronome without driving force and retards as follow:

  \begin{equation}
   \label{metronome}
\frac{d^{2}\theta}{dt^{2}}+\frac{mr_{cm}g}
{I}sin\theta=0
  \end{equation}
The angular momentum increment could be shifted into this form:
$$ p_{0}=\int_{-\infty}^{+\infty} p_{0}\delta (\theta)d\theta $$

Where $\delta(\theta)$ is Dirac's delta function.

Take Fourier Transform of $\delta (\theta)$:

  \begin{equation} \label{fourier}
    C_{\omega}=\frac{1}{2\pi}\int_{-\infty}
    ^{+\infty} \delta(\theta)e^{-i\omega \theta}d\omega=\frac{1}{2\pi}
  \end{equation}

Thus $\delta(\theta)$ becomes:
$$\delta(\theta)=\lim_{K\to +\infty}\frac{1}{\pi}\frac{sin K\theta}{\theta}$$

Then take the Taylor's expansion of $sin K\theta$:
$$\delta(\theta)=\lim_{K\to +\infty}\frac{aKp_{0}}{\pi}[1-(K\theta)^{2}
],a=\frac{3}{4}\pi$$

And due to the uniform convergence of the term $\frac{K}{\pi}[1-(K\theta)^{2}]$,$P_{0}$
becomes:

$$P_{0}=\int_{-\infty}^{+\infty} \lim_{K\to +\infty} \frac{aKp_{0}}{\pi} [1-(K\theta)^2]d\theta
=\lim_{K\to +\infty} \int_{-\infty}^{+\infty} \frac{aKp_{0}}{\pi} [1-(K\theta)^2]d\theta
=\lim_{K\to +\infty} \int_{-\infty}^{+\infty} \frac{aKp_{0}}{\pi} [1-(K\theta)^2]\dot{\theta}
dt$$

If we integrate with the substitute method, take $\theta=K\theta$ as the integral variable, then the upper and lower limit of integral becomes $\theta_{0}$, which is the maximum of the angle displacement during each half period. Then $p_{0}$ becomes:

\begin{equation}
\label{aaa}
p_{0}= \int_{-\theta_{0}}^{+\theta_{0}} \frac{ap_{0}}{\pi} [1-(\theta)^2]\dot{\theta}
dt
\end{equation}

Then take the derivatives of Eq(3) and it becomes the Van der Pol terms $\frac{ap_{0}}{\pi} [1-(\theta)^2]\dot{\theta}$ as the driving force. Take $\epsilon = \frac{ap_{0}}{\pi}$ as the momentum change after half period.

Then the motion of a single metronome can be narrated as follow:

  \begin{equation} \label{basic}
\frac{d^{2}\theta}{dt^{2}}+
\epsilon[\theta^{2}-1]\dot{\theta}
+\frac{mr_{cm}g}
{I}sin\theta=0
  \end{equation}

In addition, we explain this in a physical way. If we consider the augment of angular
displacement so small that it’s proportional to $\epsilon$ during every tick, the consecutive momentum change when $\theta=0$ could be deemed as a discrete one which is taken by the driving force step by step. What we did before is to reverse the process aforesaid.

Based on the equation aforesaid, all the parameters could be measured directly except the Van der Pol parameter $\epsilon$, thus we find the best fit to determine ε after directly measuring other parameters.

\begin{figure}[htbp]
  \centering
  \includegraphics[width=0.7\textwidth]{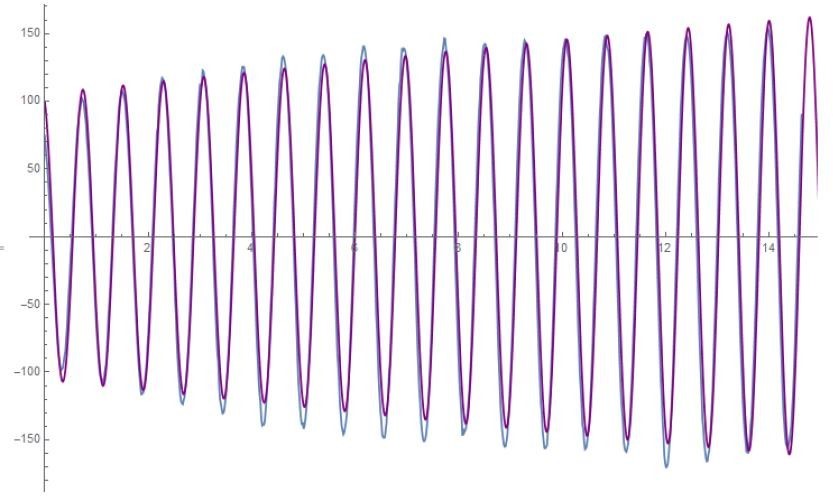}
  \caption{Contrast of theoretical and experimental curve of single metronome when the speed is 152bpm, where the purple one is theoretical and blue one is experimental.} \label{pngsample}
\end{figure}

Finally, through fitting powered by Mathematica, we get the specific volume of $\epsilon=0.01$, and the frequency $\omega^{2}=65.8$ when the speed is 152 bpm.

\subsection{Equation of several metronomes systems}

Consider the two metronomes placed on a plank which is on two cans that can roll freely without friction. This system has three degrees of freedom: two angular displacement $\theta$ and $\phi$ of the two metronomes, the coordinate of the plank $x$.

The kinetic energy of the system can be written as:
$$T=\frac{1}{2}(M\dot{x}^{2}+I_{cm}\dot{\theta}
^{2}+I_{cm}\dot{\phi}^{2})+\frac{1}{2}
m[(\dot{x}^{2}+r_{cm}cos\theta \dot{\theta})^{2}+(\dot{x}^{2}+r_{cm}cos\phi
\dot{\phi})^{2}+(r_{cm}sin\theta \dot{\theta})^{2}+(r_{cm}cos\phi \dot{\phi})^{2}]$$

Where $M$ is the total mass of the shell of the metronomes and the plank, $I_{cm}$ the momentum of inertia taking the center of mass as a reference, and $r_{cm}$ is the distance from the center of mass to the axis.

According  to  the  parallel  axis  theorem,  we  have $I_{cm}=I-mr_{cm}^{2}$, where $I$ is the momentum of inertia taking the axis as a reference.

Then the kinetic energy could be simplified as:
$$T=\frac{1}{2}(M+2m)\dot{x}^{2}+
mr_{cm}\dot{x}(cos\theta \dot{\theta}+cos\phi \dot{\phi})
+\frac{1}{2}I(\dot{\theta}^{2}+\dot{\phi}
^{2})$$

The potential could be written as:
$$V=mgr_{cm}(1-cos\theta)+mgr_{cm}(1-cos\phi)$$

The Lagrangian is:
$$L=T-V=\frac{1}{2}(M+2m)\dot{x}^{2}+
mr_{cm}\dot{x}(cos\theta \dot{\theta}+cos\phi \dot{\phi})
+\frac{1}{2}I(\dot{\theta}^{2}+\dot{\phi}
^{2})-mgr_{cm}(1-cos\theta)+mgr_{cm}(1-cos\phi)$$

After taking the term into Euler-Lagrange function, we get the equation without driving force of the system.

\begin{equation}
\frac{d^{2}\theta}{dt^{2}}+\frac{mr_{cm}
cos\theta}{I}sin\theta+\frac{mr_{cm}cos\theta}
{I}\frac{d^{2}x}{dt^{2}}=0
\end{equation}

\begin{equation}
\frac{d^{2}\phi}{dt^{2}}+\frac{mr_{cm}
cos\phi}{I}sin\phi+\frac{mr_{cm}cos\phi}
{I}\frac{d^{2}x}{dt^{2}}=0
\end{equation}

\begin{equation}
\label{eq1}
(M+2m)\frac{d^{2}x}{dt^{2}}+mr_{cm}[
[\frac{d^{2}\phi}{dt^{2}}cos\phi-(
\frac{d\phi}{dt})^{2}sin\phi]+[
\frac{d^{2}\theta}{dt^{2}}cos\theta-
(\frac{d\theta}{dt})^{2}sin\theta]]=0
\end{equation}

Considering Eq(7), we noticed that $x$ is the ignorable coordinate, which indicates that the system has a conservative law of momentum in the direction of $x$.

Thus, we take moderate coordinate as follow to simplify the equation.

$$x_{cm}=\frac{Mx+mx_{1}+mx_{2}}{M+2m},
where x_{1}=x+a+r_{cm}sin\theta,x_{2}=x+2a+r_{cm}sin\phi$$

Where $a$ is the distance between two metronomes. If we choose appropriate
coordinate system, we could eliminate the term “$a$” in the equation above.

we get $x=-\frac{m}{M+2m}r_{cm}(sin\theta+sin\phi)$

Then take the driving force with the Van der Pol terms into our account. The equation becomes:

\begin{equation}
\frac{d^{2}\theta}{dt^{2}}+
\epsilon[\theta^{2}-1]\dot{\theta}
+\frac{mr_{cm}g}
{I}sin\theta+\frac{mr_{cm}cos\theta}
{I}\frac{d^{2}x}{dt^{2}}=0
\end{equation}

\begin{equation}
\frac{d^{2}\phi}{dt^{2}}+
\epsilon[\phi^{2}-1]\dot{\phi}
+\frac{mr_{cm}g}
{I}sin\phi+\frac{mr_{cm}cos\phi}
{I}\frac{d^{2}x}{dt^{2}}=0
\end{equation}

Thus, we have got the oscillating equation of the system of two identical metronomes on a bob. For small $\epsilon$, this term will produce stable oscillations with an amplitude of approximately $2\theta_{0}$ in the isolated oscillator. As is apparent, when $\theta>\theta_{0}$, the effect of Van der Pol terms could be deemed as damping force, and when $\theta<\theta_{0}$, it serves as a driving force.

In addition, to get a more thorough inspect into the entire system, we take the subsequent transform to simplify the equation.

We denote that
$$\beta=\frac{mr_{cm}}{M+2m}\frac{mr_{cm}}{I}$$

as the coefficient parameter which narrate the extent of coupling, and then nondimensionalize (take $t=\omega t_{0}$) the equation to get the final version.

\begin{equation}
\frac{d^{2}\theta}{dt^{2}}+
\epsilon[\theta^{2}-1]\dot{\theta}
+sin\theta+\beta cos\theta \frac{d^{2}}{dt^{2}}(sin\theta+sin\phi)=0
\end{equation}

\begin{equation}
\frac{d^{2}\phi}{dt^{2}}+
\epsilon[\phi^{2}-1]\dot{\phi}
+sin\phi+\beta cos\theta \frac{d^{2}}{dt^{2}}(sin\theta+sin\phi)=0
\end{equation}

\section{Analytical Approximate Method.}

Subsequently, we use two different approximate method to analyze the secular deed of the nonlinear system, KBM method which enable us to write the initial condition into our equation and multi-scale method which aims at eliminating secular terms that is impossible from a physical point of view. Given that the initial condition would tremendously impact the synchronization time, we utilize KBM method to investigate the synchronization time according to various parameters.

\subsection{Multi-Scale Method}

As the numerical solution and experimental phenomena aforesaid indicate, we aim at finding the solution in the type of $\theta=acos(\omega t+\psi)$. In the prototype of the solution, $a$ and $\psi$ are deemed as the variables that change in the long scale of time, thus defining them as a function of $T=\epsilon t$ , where $\epsilon$ is so small that it can pander for the prerequisite that Van der Pol oscillator requires.

This method is called multi-scale perturbation, which could be utilized to cope with the nonlinear equation system with different scale of length of time with the assumption to separate time into different time scale, and assume different times are independent, namely $t$, $T_{1}=\epsilon t$, $T_{2}=\epsilon^{2} t$ and so on. This perturbative method is designed to avoid secular terms, and thus yielding an approximate solution that models the true solution for all time.

After substitute the dimensionless time into our equation, we get the simplified equation as follow:

\begin{equation}
\frac{d^{2}\theta}{dt^{2}}+
\mu[\theta^{2}-1]\dot{\theta}
+(1+\Delta)sin\theta+\beta cos\theta \frac{d^{2}}{dt^{2}}(sin\theta+sin\phi)=0
\end{equation}

\begin{equation}
\frac{d^{2}\phi}{dt^{2}}+
\mu[\phi^{2}-1]\dot{\phi}
+(1-\Delta)sin\phi+\beta cos\theta \frac{d^{2}}{dt^{2}}(sin\theta+sin\phi)=0
\end{equation}

When taking the latent discrepancy of frequency between two metronomes into our consideration, we define two dependent parameter to simplify our equation:

$$\Delta=\frac{\omega_{1}-\omega{2}}{\omega}$$
$$\beta=\frac{mr_{cm}}{M+2m}\frac{mr_{cm}}{I}$$

And taking the zero-order solution as follow:

$$\theta=A\theta_{0}cos(\tau+\varphi)$$
$$\phi=B\phi_{0}cos(\tau+\xi)$$
$$\psi=\varphi-\xi$$

Taking these terms into the system equation, to avoid secular terms appears in our equation to result in the immoderately energy augment due to the resonation (which is impossible from s physics point of view), the coefficient before the resonant terms must vanish, thus the evolution equation of amplitude and phase versus time could be written as

$$\frac{d\psi}{dt}=\frac{1}{8}[-3\gamma(
A^{2}-B^{2}+8\Delta+4\beta(\frac{B}{A}-\frac{A}{B})
cos\psi)]$$
$$\frac{dA}{d\tau}=\frac{1}{8}[\mu A(4-A^{2})+d\beta Bsin\psi]$$
$$\frac{dB}{d\tau}=\frac{1}{8}[\mu B(4-B^{2})+d\beta Asin\psi]$$

In addition, we define two new dependent variables and take another transformation as follow to decouple the equation above to get the fixed point:

$$r=\frac{A^{2}+B^{2}}{4},s=(\frac{B}{A}-\frac{A}{B})$$

Finally, we get the decoupled equation the would determine the fixed point.

$$\frac{dr}{d\tau}=\mu r[1-(\frac{s^{2}+2}{s^{2}+4})r]$$
$$\frac{ds}{d\tau}=-\frac{1}{2}[\mu sr+\beta(s^{2}+4)sin\psi]$$
$$\frac{d\psi}{d\tau}=\frac{1}{2}[(
\frac{3\gamma r}{\sqrt{s^{2}+4}}+\beta cos\psi)s+2\Delta]$$

Observing the first evolution equation, because both the $r$ and $d$ are dependent term and non-negative, then we get the attractive fixed point $r^{*}$.

$$r\to \frac{s^{2}+4}{s^{2}+2}$$

Then we substitute the fixed point into the other two equations, and get two fixed point with one attractive and the other one repulsive. The attractive fixed point is that

$$s\to \frac{\mu}{2\beta sin\psi}[-1+\sqrt{1-2(2\beta sin\psi/\mu)^{2}}]\approx -\frac{2\beta}{\mu}sin\psi$$

Considering that $\mu$, $\beta$ and $\gamma$ are all small parameters, the equation above could be simplified further. In-phase fixed point phase difference is small for small $\Delta$, so the fixed point of $s$ can be simplified as

$$s^{*}\approx -\frac{2\beta}{\mu}sin\psi$$

And take this simplified fixed-point value into the phase difference evolution equation, we get a phase difference given by attractive fixed point, from which we could educe the prerequisite of synchronization.

$$\psi \approx arcsin(\frac{\mu \Delta}{\beta(3\gamma+\beta)}$$

These analyses only work when $\Delta$ is so small that the condition of these approximation makes sense. For larger $\Delta$, the phase difference $\psi$ becomes larger and the approximation breaks down. At large $\Delta$ ,the threshold value of $\psi$, where synchronization is no longer possible, agrees with the value where $s^{*}$ becomes imaginary.

In addition, as we take the categories of synchronization, the sign of $cos\psi$ could be utilized as a resort. $cos\psi>0$ denotes the in-phase synchronization, and vice versa.

What’s more, the $cos\psi$ could also embody the stability of the system, which demonstrates that if $cos\psi>0$, the solution is stable, and if $cos\psi<0$, which means temporary anti-phase synchronization, it would eventually be attracted to the stable solution of in-phase synchronization, which correspond with the numerical solution we mentioned above.

If the pertinent parameters are so large that the phase difference evolution equation has no solution, the synchronization can never be accessed.

\subsection{KBM Method}

KBM method is an efficient and feasible approximate approach which could provide a linearization of nonlinear system, where Taylor’s expansion seems more plausible and reliable than in nonlinear system because nonlinear term must have more complex impact on variables. Base on some restriction and contrast between the approximate solution and real solution, KBM method would provide a high accurate linearized system which share the same secular motion and most details with the former nonlinear system.

Consider a general nonlinear system as follow:

\begin{equation}
\ddot{x}+p_{0}^{2}x=\epsilon f(x,\dot{x})
\end{equation}

Then, we write the first order approximate solution as follow:

$$x=acos\psi+\epsilon x_{1}(a,\psi)$$
$$\frac{da}{dt}=\epsilon A_{1}(a)$$
$$\frac{d\psi}{dt}=p_{0}+\epsilon p_{1}(a)$$

And as for its real solution, we could deem $a$ and $\psi$ as a function of time, and based on what the numerical results and our former discussion, we could write the pertinent parameters as:

$$\psi=p_{0}t+\phi(t)$$

The restriction aforesaid that ascertain the accuracy of KBM method is that after our approximation the form of term of speed must attain the same as the circumstance where $\epsilon=0$, which is

$$\frac{dx}{dt}=-ap_{0}sin\psi$$

Thus, we write down the real solution and take the first derivative of the displacement and get

\begin{equation}
\frac{dx}{dt}=-ap_{0}sin\psi+\frac{da}{dt}cos\psi
-a\frac{d\phi}{dt}sin\psi
\end{equation}

By contrast with the former expression of speed we get one of our approximate equation:

$$\frac{da}{dt}cos\psi-a\frac{d\phi}{dt}sin\psi=0$$

In addition, we could get the acceleration term of the real solution as follow

\begin{equation}
\frac{d^{2}x}{dt^{2}}=-ap_{0}cos\psi-
\frac{da}{dt}p_{0}sin\psi-ap_{0}\frac{d\phi}{dt}cos\psi
\end{equation}

Taking Eq(14) and Eq(15) back into Eq(13) we get the second relationship between the approximate solution and real solution:

$$\frac{da}{dt}p_{0}sin\psi+ap_{0}
\frac{d\phi}{dt}cos\psi=-\epsilon f(acos\psi,-ap_{0}sin\psi)$$

Given that $a$ and $\psi$ is considered as a function of time, we could get the solution of $a$ and $\psi$ through the two relationship we got before.

$$\frac{da}{dt}=-\frac{\epsilon}{p_{0}}
f(acos\psi,-ap_{0}sin\psi)sin\psi$$
$$\frac{d\phi}{dt}=-\frac{\epsilon}{ap_{0}}
f(acos\psi,-ap_{0}sin\psi)cos\psi$$

Until now, what we have done is only the contrast between the approximate solution and real solution without any approximation. What we will do next is substitute the average value in a whole period into the equation above to make terms in the right of the equation become a function that is only about the independent variable $\psi$. After this approximation, the equation become

$$\frac{da}{dt}=-\frac{\epsilon}{2\pi p_{0}}\int_{0}^{2\pi}f(acos\psi,-ap_{0}sin\psi)
sin\psi d\psi$$
$$\frac{d\phi}{dt}=-\frac{\epsilon}{2\pi ap_{0}}
\int_{0}^{2\pi}f(acos\psi,-ap_{0}sin\psi)
cos\psi d\psi$$

Subsequently, we will derive the equivalent linearized system of our primal equation.
We denote several dependent variables to simplify our derivation:

$$\frac{d\psi}{dt}=p_{e}(a)$$
$$p_{e}^{2}(a)=p_{0}^{2}-\frac{\epsilon}{\pi ap_{0}}
\int_{0}^{2\pi}f(acos\psi,-ap_{0}sin\psi)
cos\psi d\psi+o(\epsilon^{2})$$

Note that

$$\lambda_{e}^{2}(a)=\frac{\epsilon}{\pi ap_{0}}
\int_{0}^{2\pi}f(acos\psi,-ap_{0}sin\psi)
sin\psi d\psi$$
$$K_{e}(a)=K-\frac{\epsilon}{ap_{0}}
\int_{0}^{2\pi}f(acos\psi,-ap_{0}sin\psi)
cos\psi d\psi$$
where K is the constant of damping.

Then we take the first and second order derivatives of the displacement and take them into the primal equation after neglecting the seconder order terms, we get our equivalent linearized system as follow:

$$\frac{d^{2}x}{dt^{2}}+\frac{K_{e}(a)}{m}
x+\frac{\lambda_{e}(a)}{m}\frac{dx}{dt}=0$$

When we take the model of metronomes system into our approximate system discussed before, the equivalent linearized system become:

\begin{equation}
\frac{d^{2}\theta}{dt^{2}}+\frac{mr_{cm}g}{I}\theta-
\epsilon(1-\frac{1}{1+(-1+\frac{4}{a_{0}^{2}})e^{-\epsilon t}})\frac{d\theta}{dt}-\frac{mr_{cm}}{I}\frac{d^{2}}{dt^{2}}(\theta+
\phi)=0
\end{equation}

\begin{equation}
\frac{d^{2}\phi}{dt^{2}}+\frac{mr_{cm}g}{I}\phi-
\epsilon(1-\frac{1}{1+(-1+\frac{4}{b_{0}^{2}})e^{-\epsilon t}})\frac{d\phi}{dt}-\frac{mr_{cm}}{I}\frac{d^{2}}{dt^{2}}(\theta+
\phi)=0
\end{equation}

Where $a_{0}$, $b_{0}$ are the initial angle displacement of the two metronomes separately.
The reason why we choose KBM method from other myriad of approximate methods in nonlinear dynamics is that this method make it possible to take the initial condition into the equation so that we could have a direct research about how the initial condition influence the whole system especially when it comes to problems about the independent variable $t$.

What’s more, we come to the verification of the accuracy of our approximate equation using the contrast of the numerical solution with the approximate solution. In both the two figures, the orange curve is the solution of primal equation, and the purple curve is the approximate solution. From our contrast, the accuracy of the linearization method could be demonstrated outstanding enough.

\begin{figure}[htbp]
  \centering
  \includegraphics[width=0.7\textwidth]{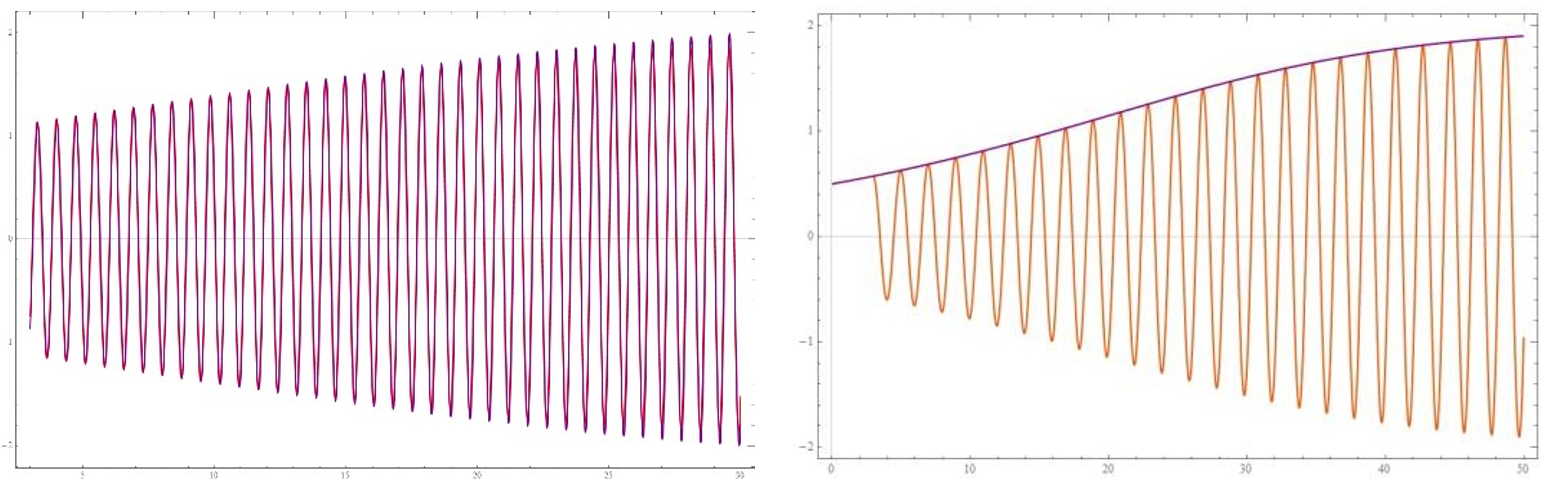}
  \caption{The contrast between primal system and approximate system of two coupled metronomes.} \label{pngsample}
\end{figure}

Then, we will get back to the approximate system to formulate the relationship between synchronization time and the initial condition. However, due to the discrepancy between these two initial condition, it’s impossible to make direct calculation between the two equation to decouple, so we take the Taylor expansion of the linearized system and only leave the first order terms. Then, due to the fact that when synchronization is achieved, the term $(\dot{\theta}-\dot{\phi})$ must be tend to zero, which means that the terms before $(\dot{\theta}+\dot{\phi})$ must be small enough so that it will not have tremendous impact on the whole system, so we could get the approximate formula of synchronization time as follow

$$\Delta a\frac{8\epsilon e^{-\epsilon t}}{\bar{a}^{3}(1+(-1+\frac{4}{\bar{a}^{2}})e^{-
\epsilon t})^{2}}<\delta_{0}$$

Where $\bar{a}$ denotes the average of two different initial angle displacement, and $\Delta a$ denotes the discrepancy between these two initial values. To get the specific synchronization time, $\delta_{0}$ must be determined through fitting of experimental results. In our experiment, we use the first set of statistic to get the specific volume of $\delta_{0}=0.00476$ , and then the curve of synchronization time versus initial angle displacement could be plotted using numerical method.

\begin{figure}[htbp]
  \centering
  \includegraphics[width=0.7\textwidth]{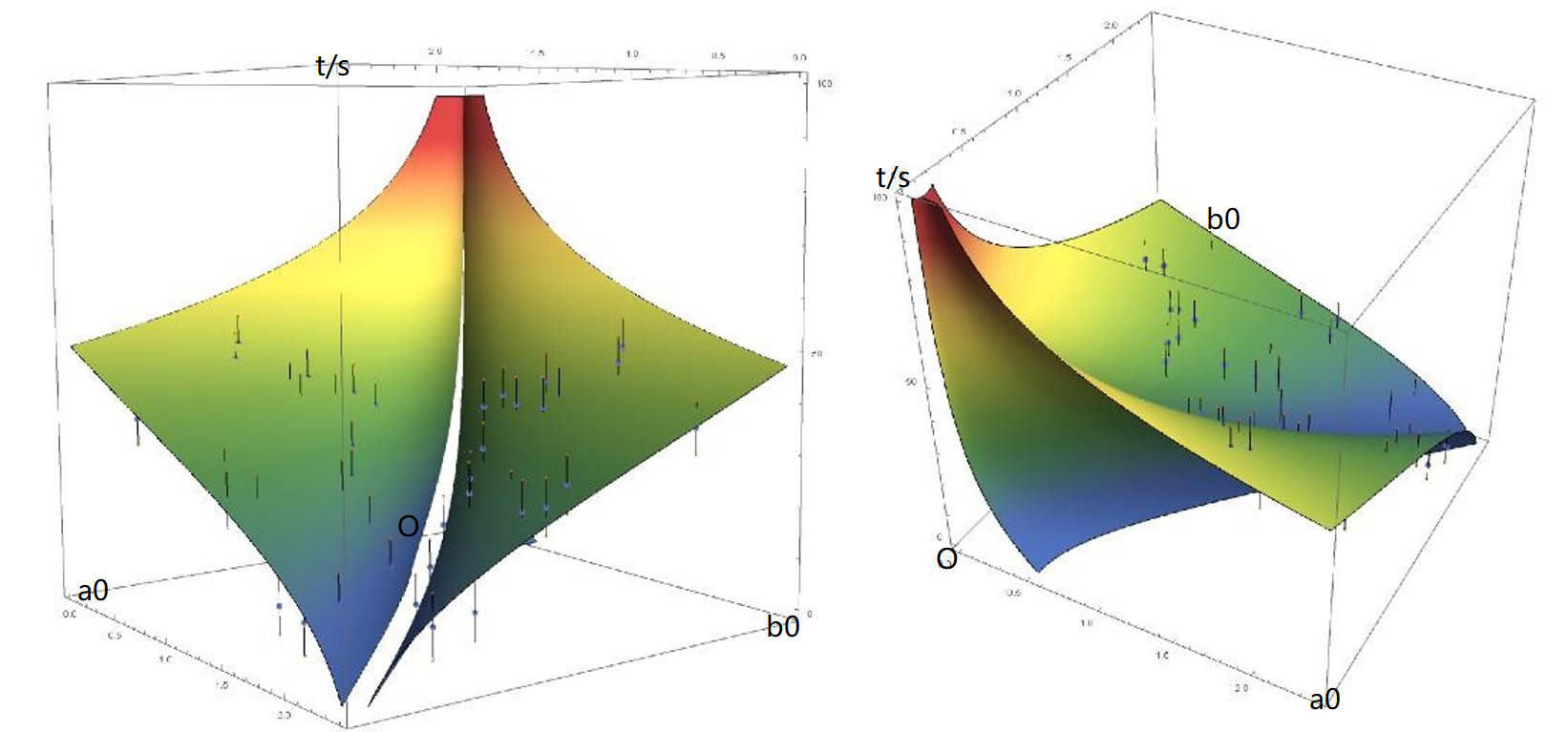}
  \caption{Theoretical and experimental curve of synchronization time versus initial angle displacement, where curved surface is a plot of synchronization versus the sum of initial angle displacement and the discrepancy of that.} \label{pngsample}
\end{figure}

From the curved surface, one of the basic conclusion of our experiment could be derived is that the less the difference between the two initial conditions, the less time it would take to attain synchronization, which is obvious and trifle. What is non-trifle is that the bigger the sum of these two initial angle displacement, the less time it will take.

In addition, what our present theoretical deduction could not account for now is that the synchronization time has nothing to do with their frequency (actually we have neglected the discrepancy of natural frequency between these two metronomes). However, in our experiment, we observed that the frequency does impact the synchronization time to a tremendous extent. Hence, we consider ameliorating our theoretical formula according to the restriction of KBM method aforesaid.

Given that when synchronization is attained, it is assumed that $(\theta-\phi)<\delta_{2}$. According to the KBM approximation restriction given before, we urge that after our approximation the form of term of speed must attain the same as the circumstance where $\epsilon=0$.

$$\frac{dx}{dt}=-ap_{0}sin\psi$$
$$sin\psi_{1}-sin\psi_{2}<\delta_{3}$$

In a similar way, the parameter before term $(\theta+\psi)$ must be small enough, which is

$$\Delta a\frac{8\epsilon e^{-\epsilon t}}{\bar{a}^{3}p(1+(-1+\frac{4}{\bar{a}^{2}})e^{-
\epsilon t})^{2}}<\delta_{0}$$

Where $p$ is the natural frequency of the metronome. From the formula above, we could conclude that the bigger the natural frequency is, the less time it will take to synchronize, which correspond to our experiment. In a similar way, we could get the specific value of $\delta_{4}$ through the first set of statistic where $\delta_{4}=0.0048$.

\begin{figure}[htbp]
  \centering
  \includegraphics[width=0.7\textwidth]{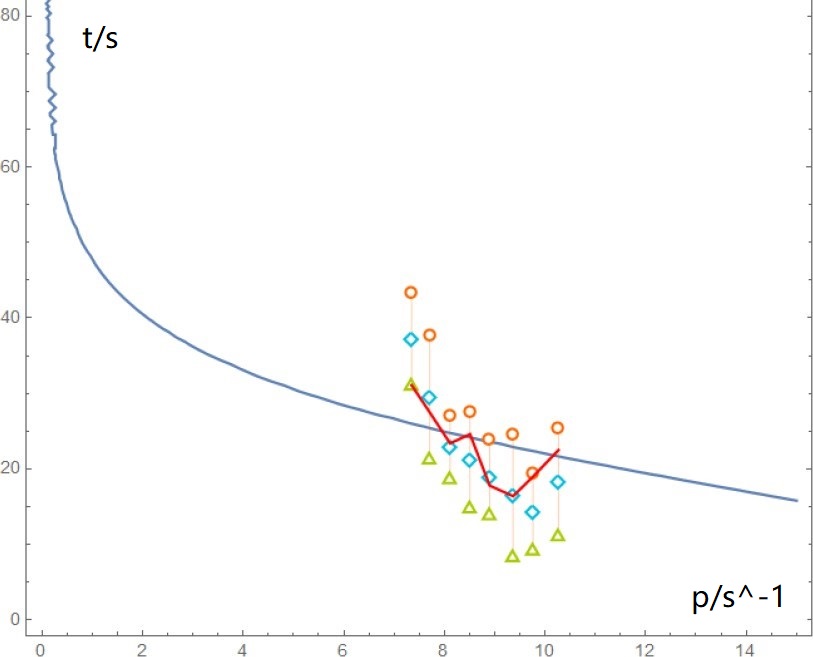}
  \caption{The ameliorated theoretical curve of synchronization time versus natural frequency} \label{pngsample}
\end{figure}

In the curve above, the blue line is the theoretical curve where the average of initial phase $a$ and the discrepancy between two initial condition $\Delta a$ are fixed. The blue point is the experimental result and the red line is the actual theoretical result where $\Delta a$ and $a$ are both their real value. Actually, it is too difficult to control the same initial condition between two experiments, so we compromise and use this method to have a legible overview of how the frequency impact on the synchronization time.

From both experimental and theoretical result, we could conclude that the frequency does have indispensable impact on the synchronization time. And the larger the frequency is, the less time it will take to synchronize.

\subsection{Numerical Results}

Although, as to a nonlinear system, giving out its specific analytic equation of their attractive area seems impossible, we could still use numerical result to inspect the impact of initial condition and the friction parameter $\beta$ on the type of synchronization. Our numerical simulation result is given as follow:

\begin{figure}[htbp]
  \centering
  \includegraphics[width=0.7\textwidth]{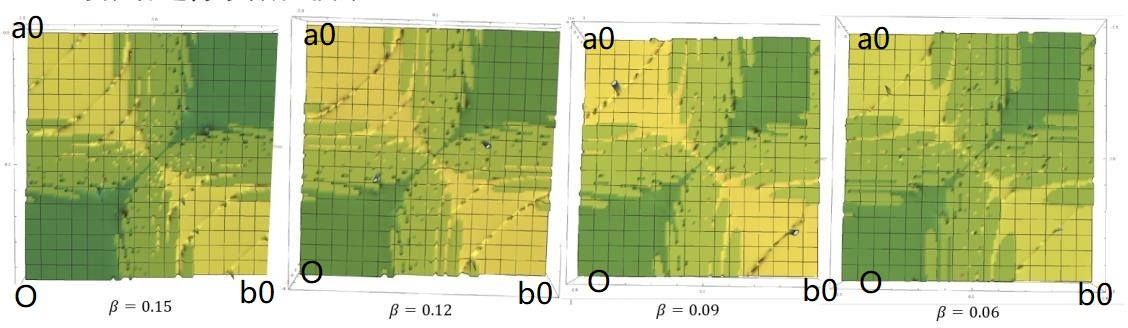} \label{pngsample}
\end{figure}

Where the dark green point denotes anti-phase synchronization and the yellow point denotes in-phase synchronization. If these two metronomes are not yet synchronize, the status will be denoted by shallow green point.

From the numerical result above, it can be concluded that a larger $\beta$ will result in less available anti-phase synchronization, which in return could trigger more in-phase synchronization.

However, these result is not compatible now to be verified by experiment because that the specific border of the attractive area would not be measured accurately. For instance, when the initial angle displacement was taken deeply into one area such as in-phase synchronization attractive area, we could know for sure that the type of synchronization must be the in-phase one. When the test point is near the border, every small perturbation may cause the difference in the experimental result of type of synchronization. Hence, we could only investigate the type of synchronization through numerical simulation.

\section{Conclusion and Discussions}

Synchronization phenomenon are common and inevitable especially in the area of signal or ecology. Researchers of this paper have proved the compatibility of Van der Pol equation in the nonlinear system with discrete small driving force or damping.

In addition, researchers have discussed a specific example of synchronization in the metronomes system, and have derived the equation of the system, provided the multi- time scale method to dispose with a general sort of nonlinear system, with the preponderance of the spurn of secular terms and the accuracy every rank of approximation. Also, KBM method was utilized to take the initial condition into the primal equation to get the synchronization time. Researchers have a thorough investigation into the parameters that could impact on the synchronization time and the type of synchronization.

Moreover, researchers investigate how the initial condition impact the evolution of the equation, and conclude that the less the difference between the two initial conditions, the less time it would take to attain synchronization, which is obvious and trifle. What is non- trifle is that the bigger the sum of these two initial angle displacement, the less time it will take. In addition, researchers provided a amelioration of inherent frequency of the coupled system and get the result that how the inherent frequency influence the synchronization time and conclude that the bigger the natural frequency is, the less time it will take to synchronize.

What’s more, from our numerical result, to verify the influence between initial condition and the type of synchronization, it can be concluded that a larger β will result in less available anti-phase synchronization, which in return could trigger more in-phase synchronization.

For further study, this model could be dilated into the synchronization of several metronomes with identical or non-identical inherent frequency, with or without even distribution of their position. When the number of metronomes increase, the process of getting to synchronize could also be taken into the researchers account for a thorough consciousness of the synchronization of complex dynamics system.

%%%%%%%%%%%%%%%%%%%%%%%%%%%%%%%%%%%%%%%%%%%%%%%%%%%%%%%%%%%%%%%%
%  References
%%%%%%%%%%%%%%%%%%%%%%%%%%%%%%%%%%%%%%%%%%%%%%%%%%%%%%%%%%%%%%%%

\end{document}